\documentclass[twocolumn,prb,aps,preprintnumbers,longbibliography, balancelastpage, superscriptaddress]{revtex4-2}
\usepackage{amsmath,amssymb,bm,graphicx,color,gensymb,bbold,appendix,xspace}
\usepackage[colorlinks=true, citecolor={blue!80!black}, urlcolor={blue!50!black}, linkcolor = {blue!80!black}]{hyperref}
\usepackage[utf8x]{inputenc}

\usepackage{xcolor}

\newcommand{\KMSO}{K$_2$Mn(SeO$_3$)$_2$\xspace}
\newcommand{\KCSO}{K$_2$Co(SeO$_3$)$_2$\xspace}
\newcommand{\KNSO}{K$_2$Ni(SeO$_3$)$_2$\xspace}
\newcommand{\BCSO}{Ba$_3$CoSb$_2$O$_9$\xspace}

\begin{document}

\title{Dynamics and thermodynamics of the $S=5/2$ almost Heisenberg triangular lattice antiferromagnet \KMSO}

\author{Mengze Zhu}
\email{zhumen@phys.ethz.ch}
\address{Laboratory for Solid State Physics, ETH Z\"{u}rich, 8093 Z\"{u}rich, Switzerland}

\author{V. Romerio}
\address{Laboratory for Solid State Physics, ETH Z\"{u}rich, 8093 Z\"{u}rich, Switzerland}

\author{D. Moser}
\address{Laboratory for Solid State Physics, ETH Z\"{u}rich, 8093 Z\"{u}rich, Switzerland}

\author{K.~Yu.~Povarov}
\address{Dresden High Magnetic Field Laboratory (HLD-EMFL) and W\"urzburg-Dresden Cluster of Excellence ctd.qmat, Helmholtz-Zentrum Dresden-Rossendorf, 01328 Dresden, Germany}

\author{R. Sibille}
\address{PSI Center for Neutron and Muon Sciences, 5232 Villigen PSI, Switzerland}

\author{R. Wawrzy\'{n}czak}
\address{PSI Center for Neutron and Muon Sciences, 5232 Villigen PSI, Switzerland}

\author{Z. Yan}
\address{Laboratory for Solid State Physics, ETH Z\"{u}rich, 8093 Z\"{u}rich, Switzerland}

\author{S. Gvasaliya}
\address{Laboratory for Solid State Physics, ETH Z\"{u}rich, 8093 Z\"{u}rich, Switzerland}

\author{A. L. Chernyshev}
\address{Department of Physics and Astronomy, University of California, Irvine, California 92697, USA}

\author{A.~Zheludev}
\email{zhelud@ethz.ch; http://www.neutron.ethz.ch/}
\address{Laboratory for Solid State Physics, ETH Z\"{u}rich, 8093 Z\"{u}rich, Switzerland}

\begin{abstract}
We report calorimetric, magnetic, and neutron scattering studies on an $S = 5/2$, nearly Heisenberg triangular-lattice antiferromagnet \KMSO with weak XXZ easy-axis anisotropy. Multiple magnetic phases are identified, including a non-collinear Y phase in zero field, a field-induced collinear $m = 1/3$ magnetization plateau, and a high-field V phase. In the Y phase, the magnetic excitation spectrum exhibits both single-magnon excitations and an extended high-energy continuum. Both features are well described by non-linear spin wave theory. In the field-induced phases, complex effects of the spectrum renormalization even for large $S=5/2$ material are clearly detectable. These results underscore the essential role of magnon-magnon interactions in the dynamics of large-$S$ Heisenberg spin systems on a triangular lattice.

\end{abstract}

\date{\today}
\maketitle

\section{Introduction}
The Heisenberg spin Hamiltonian on a triangular lattice is an archetype of geometrical frustration. Even though the quantum model has semiclassical $120^\circ$ long-range order in the ground state, it has become an important starting point for numerous routes towards exotic phases such as quantum spin liquids \cite{Balents2010,ZhuWhite_PRB_2015_TriangularNNN,ZhuMaksimov_PRL_2017_YMGOmimicry,ZhuMaksimov_PRL_2018_AnisotropicTriangLattice,gallegos2024}.  Indeed, frustration of interactions and low dimensionality ensure that quantum effects remain significant. The most quantum $S=1/2$ case has been the subject of numerous recent studies. Of particular interest are anomalous features in the excitation spectrum that have been experimentally detected in such model compounds as \BCSO \cite{Itoh2017,MacdougalWilliams_PRB_2020_BaCoSbOtriangularexcitations} and Yb-based delafossites \cite{DaiZhang_PRX_2021_NaYbSespinonFS,Scheie2024,Xie2023}. In contrast to the predictions of linear spin wave theory (LSWT), a substantial spectral weight is contributed by a broad continuum, rather than by coherent single-particles excitations. The continuum extends to energies well beyond the scale set by  Heisenberg exchange constant $J$. Even the sharp modes are anomalous, with dispersion relations deviating significantly from those predicted by LSWT.

Several rather exotic theoretical scenarios have recently been put forward to explain these observations. One view is that the anomalous features may be associated with fractional spinon excitations of nearby quantum spin liquids \cite{Ghioldi2022,Bose2025}. The opposing more pedestrian interpretation is in terms of conventional magnon-magnon interactions. Indeed, the phenomenon of spontaneous magnon decay and a renormalization of magnon energies in Heisenberg triangular lattice models has been extensively investigated theoretically \cite{ZhitomirskyChernyshev_RMP_2013_DecayReview,Starykh2006,Chernyshev2006,Chernyshev2009}. 
One can expect this view to be particularly well-suited for understanding more classical higher-$S$ systems. Unfortunately, experimental work in this area remains scarce due to the lack of suitable materials. Quite a bit of work has been done on  Fe- \cite{White2013,Saha2024} and Mn-based compounds \cite{Ishii2011,Lee2014,Rawl2019,Kim2022,Mingfang2023,Biniskos2025}. In what concerns details of the excitation spectrum, by far the most extensive data are available for  Ba$_3$MnSb$_2$O$_9$ \cite{Mingfang2023}. Here the measured magnon dispersion can be qualitatively described by LSWT, but the observed linewidths are considerably broader than the instrumental resolution, a signature of magnon decay. Unfortunately, due to a limited experimental energy transfer range, the existence and strength of higher-energy continuum excitations could not be ascertained.

\begin{figure}[ht]
	\includegraphics[width=\columnwidth]{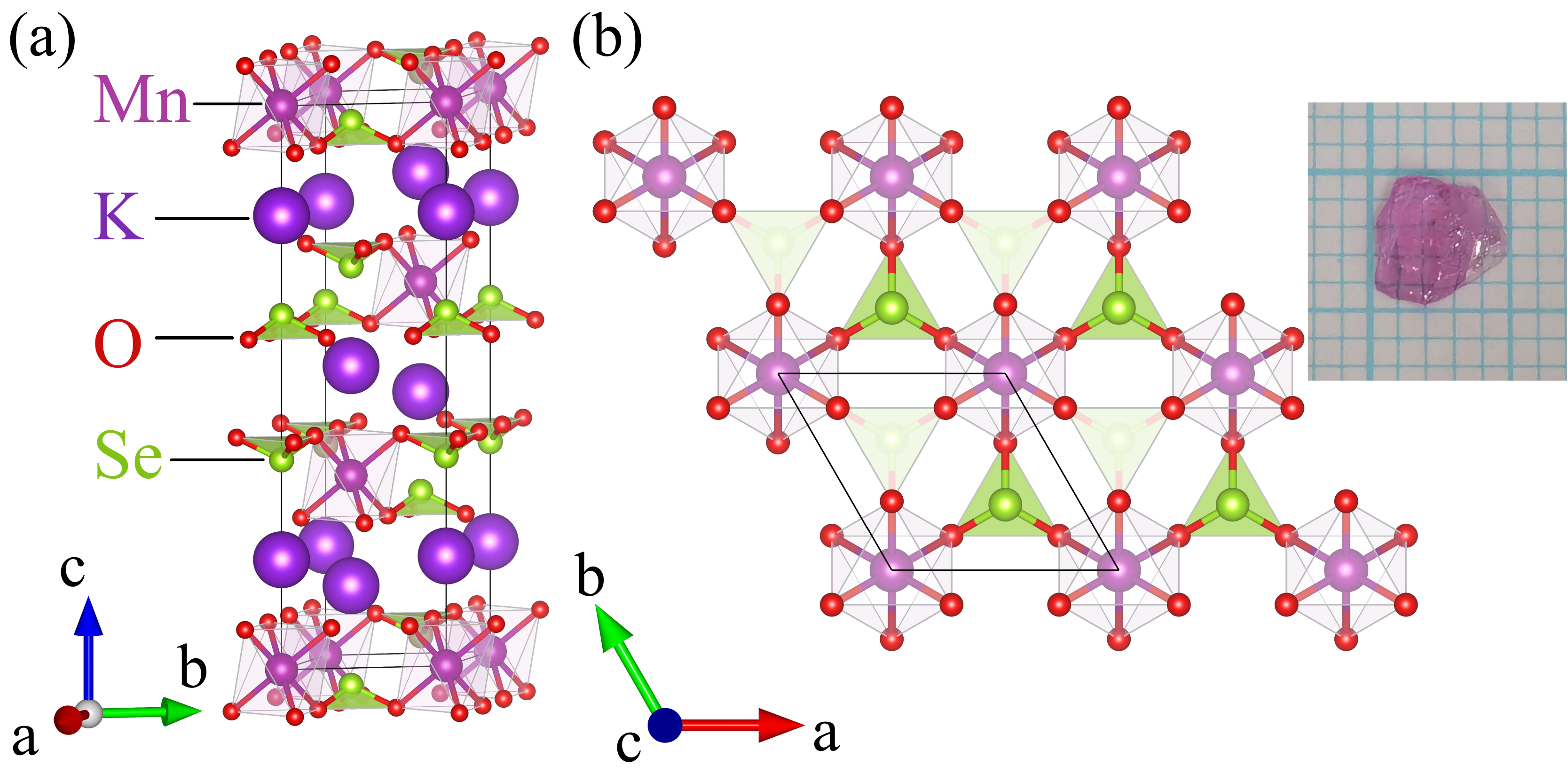}
	\caption{(a) Crystal structure of \KMSO. (b) Top view of a triangular plane consisting of MnO$_6$ octahedra connected by SeO$_3$. Inset: Photo of a \KMSO single crystal on a millimeter grid paper.}\label{fig:structure}
\end{figure}

In this paper, we report the thermodynamic and dynamical properties of a nearly Heisenberg and exceptionally two-dimensional $S = 5/2$ triangular lattice antiferromagnet \KMSO. After mapping out the magnetic phase diagram and establishing a model spin Hamiltonian for this system, we focus on the magnetic excitation spectra in each of the field-induced phases. In the non-collinear Y phase in zero field, in  addition to conventional spin waves, we observe a distinctive broad excitation continuum at higher energies, in close quantitative agreement with the analysis by the non-linear SWT calculations. The continuum is suppressed in the field-induced collinear $1/3$-magnetization plateau phase and remains much weakened in the high-field V phase. With increasing field, the energies of single-magnon modes are renormalized in a non-monotonic fashion, in agreement with the expectations from the non-linear SWT.

\begin{figure}
	\includegraphics[width=\columnwidth]{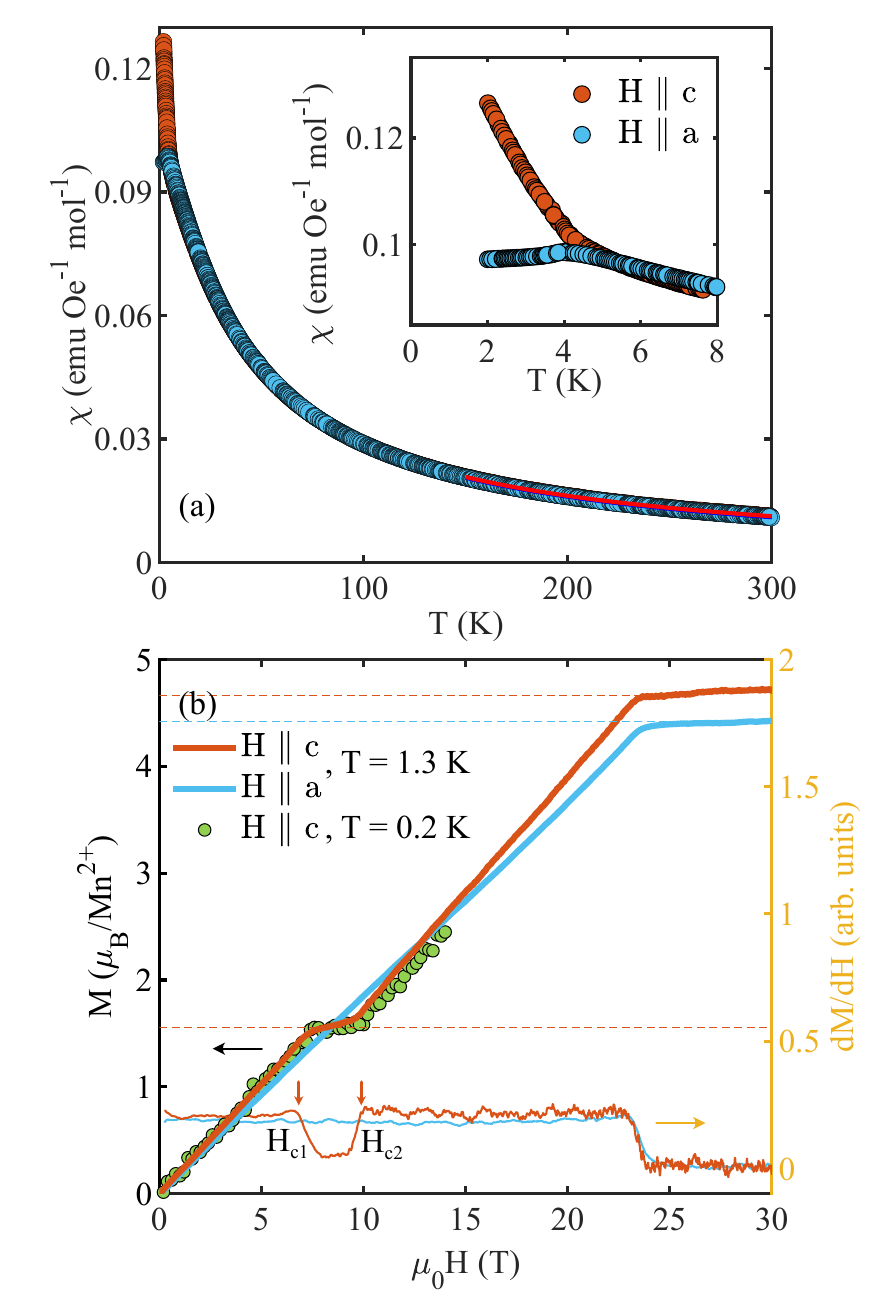}
	\caption{(a) Magnetic susceptibility as a function of temperature with $\mu_0H = 0.1$ T along the crystallographic $c$ and $a$ axis, respectively. Solid lines are the Curie-Weiss fits. Inset shows the expanded view of the low temperature regime. (b) Left axis: Magnetization as a function of magnetic fields measured using pulsed-fields at $T = 1.3$ K (solid lines) and Faraday balance magnetometry at $T = 0.2$ K (green circles). Dashed lines denote the saturation magnetization for both field orientations and 1/3 of the saturated magnetization for $H \parallel c$. Right axis: $dM/dH$ as a function of magnetic field. $H_{c1}$ and $H_{c2}$ denote the critical fields of the onset and end of the 1/3 plateau phase, respectively. }\label{fig:magnetometry}
\end{figure}

\begin{figure*}[th]
	\includegraphics[width=\textwidth]{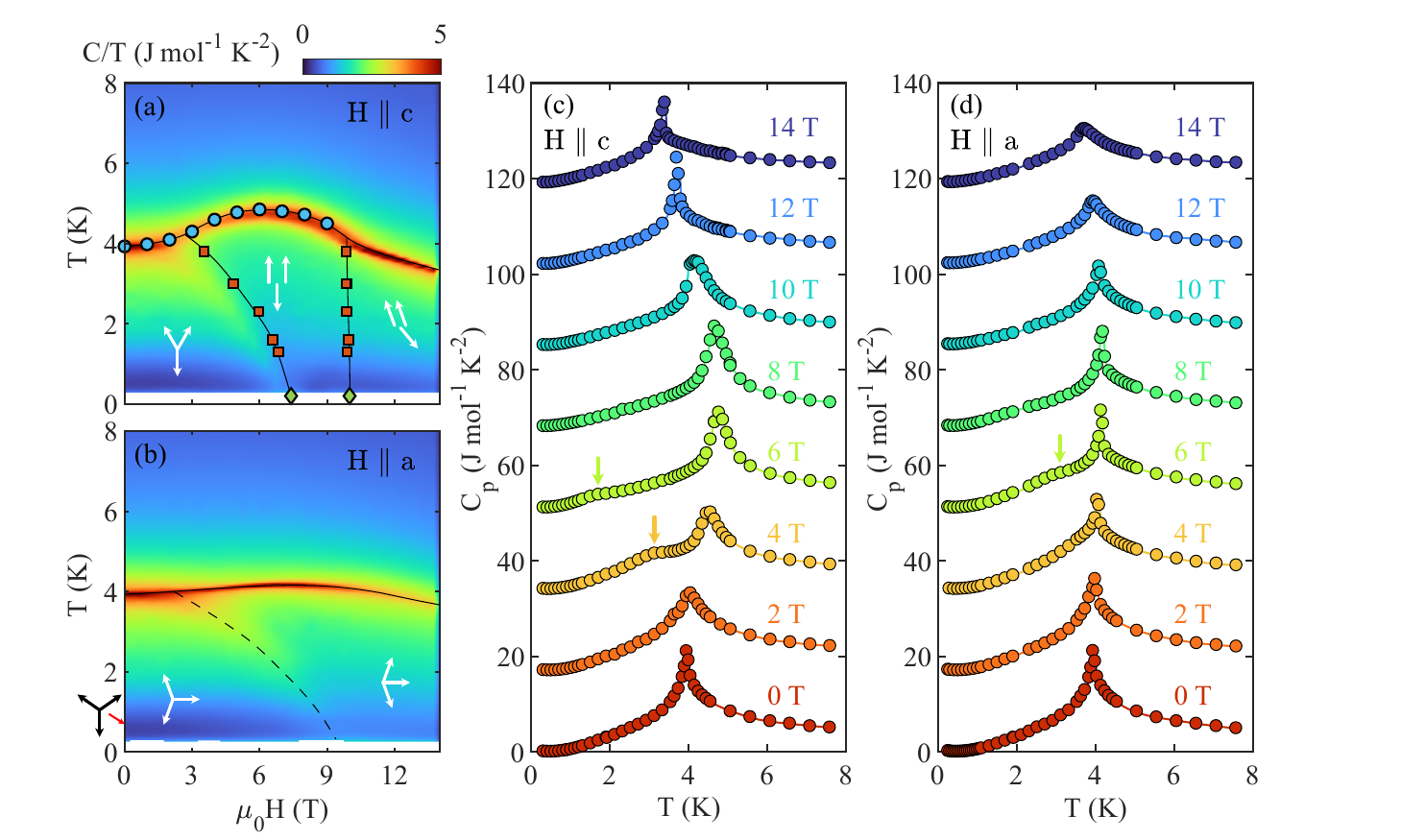}
	\caption{(a),(b) False color plots of specific heat $C_p/T$ of \KMSO measured as a function of temperature and magnetic fields along the $c$ and $a$ axis, respectively. Inset arrows illustrate the magnetic structure in each phase. Blue circles are phase boundaries determined by the neutron diffraction data. Red squares are critical fields $H_{c1}$ and $H_{c2}$ extracted from $dM/dH$ of the pulsed-field magnetization. Green diamonds are critical fields obtained from Faraday balance magnetometry. Solid lines are guides to the eye. The dash line is a crossover. (c),(d) Typical temperature dependent specific heat curves of \KMSO measured with fields along the $c$ and $a$ axis. Arrows denote the lower-temperature Y-to-uud phase transition or inverted-Y to $\Psi$ crossover as described in the text. Each curve is offset from one another by 17 J mol$^{-1}$ K$^{-1}$ for visibility. }\label{fig:specific_heat}
\end{figure*}

\section{Material and Methods}

\KMSO is member of a family of triangular lattice antiferromagnets, A$_2$M(SeO$_3$)$_2$ (A=K,Rb, M=Co,Ni,Mn), which has attracted dramatic interest recently \cite{Zhong2020,Zhu2024,Zhu2025,Chen2024, Li2023}. \KCSO is an ideal material realization of the $S$ = 1/2 XXZ Hamiltonian close to the Ising limit. It has a novel spin supersolid ground state, with anomalous excitations spectrum of strong quantum origin \cite{Zhu2024,Zhu2025,Chen2024}. In contrast, \KNSO is an $S=1$ nearly Heisenberg system with easy-plane single-ion anisotropy \cite{Li2023}. \KMSO is isostructural to \KCSO and \KNSO (Space group $R$-$3m$, $a=5.615$~\AA, $c=18.667$~\AA~\cite{Wildner1992}). Magnetism in this compound arises from the ABC-stacked Mn$^{2+}$ $S$ = 5/2 ions, which forms equilateral 2D triangular lattice separated by layers of $K^{+}$ ions, as illustrated in Figure~\ref{fig:structure}. To our knowledge, the magnetic properties and model Hamiltonian of \KMSO have not been established.

Single crystals of \KMSO were grown using similar procedures as for \KCSO \cite{Zhong2020}. The structure of the as-grown crystals was determined by a Bruker single crystal x-ray diffractometer, and is in agreement with that reported previously \cite{Wildner1992}. Magnetic susceptibility and magnetization were measured using a vibrating sample magnetometer (VSM) installed on a Quantum Design Physical Property Measurement System (PPMS) down to $T = $ 2 K and up to $\mu_{0}H = 14$ T. Magnetization at lower temperatures was measured using a custom-made Faraday balance magnetometer \cite{Blosser2020} with a dilution refrigerator insert down to $T = $ 0.2 K. The high-field magnetization measurements were performed with a coaxial pick-up coil pulsed-field magnetometer at HLD-EFML \cite{DresdenMagnetometer}. The data were calibrated to absolute units by comparing with that measured by VSM. The specific heat capacity was measured using the relaxation method by PPMS.

Neutron diffraction experiments were carried out using the lifting-counter diffractometer ZEBRA at Paul Scherrer Institut (PSI) equipped with a 10 T vertical-field cryomaget. A single crystal sample of 157 mg was used and the incident neutron wavelength was fixed at $\lambda = 2.305$ \AA. Inelastic neutron spectroscopy measurements were conducted using the multiplexing spectrometer CAMEA \cite{CAMEA2023} at PSI with an 11 T vertical field cryomagnet. The sample measured had a mass of 970 mg, and several incident energies $E_i$ = 4.87, 5, 5.13 and 5.26 meV, were used. The energy resolution was $\Delta E \approx$ 0.19 meV in FWHM at elastic scattering. For each incident energy, the data were collected at detector bank angle $2\theta=$ -40 and -44 degrees. For each $2\theta$, the sample was rotated by 180 degrees in 1 deg steps. Data reduction were performed using the MJOLNIR \cite{MJOLNIR2020} software package. In both experiments, the samples were mounted on a copper holder with the $(H,K,0)$ plane horizontal, and cooled using a dilution refrigerator. The magnetic field was applied along the crystallographic $c$ axis. LSWT calculations were performed using the SpinW package \cite{TothLake_JPCM_2015_SpinW}. Non-linear SWT results were obtained following the approach of Ref.~\cite{Mourigal2013_TAL}.

\section{Results}
\subsection{Magnetic susceptibility and magnetization}\label{sec:magnetometry}

Figure~\ref{fig:magnetometry}(a) shows the dc magnetic susceptibility $\chi_c$ and $\chi_a$ of \KMSO as a function of temperature with $\mu_0H$ = 0.1 T applied along the crystallographic $c$ and $a$ axis, respectively, after zero-field cooling. At high temperatures, the susceptibility is nearly isotropic. A fit of the data between 150$-$300 K by the Curie-Weiss law (solid lines) yields the Weiss temperatures $\Theta_{w,c}$ = $-25.7(4)$ K, $\Theta_{w,a}$ = $-29.1(3)$ K, and the effective moments $\mu_{\text{eff,c}} =$ 5.40(1) $\mu_B$, $\mu_{\text{eff,a}}$ = 5.49(1) $\mu_B$, which are very close to the expected value $g\sqrt{S(S+1)} \approx 5.9 \mu_B$ for Mn$^{2+}$ spins. The negative Weiss temperatures imply that the dominant exchange interaction is antiferromagnetic. Using the mean field result $\Theta_w = -zJ_{\text{MF}}S(S+1)/3k_B$, where the number of nearest-neighbor spins is $z$ = 6 for a 2D triangular lattice, the mean-field exchange constant between the Mn$^{2+}$ ions can be estimated as $J_{\text{MF}} \approx$ 0.13 meV. At lower temperatures, $\chi_a$ decreases below $T_N \approx 4$ K (Figure~\ref{fig:magnetometry}(a) inset), which indicates an antiferromagnetic transition. The ratio $\Theta_w/T_N \approx 7$ suggests modest frustration in the system. In contrast, $\chi_c$ continues to increase below the ordering temperature, possibly stemming from a weak ferromagnetic coupling between the triangular planes. 

Figure~\ref{fig:magnetometry}(b) shows the field dependence of magnetization measured by the pulsed-field magnetometry (solid lines) at $T = 1.3$ K, when the field sweeps up. The right axis shows the corresponding derivatives $dM/dH$. A 1/3 magnetization plateau is observed when the field is applied along the $c$ axis, which is consistent with that measured by the Faraday balance magnetometer at a lower temperature $T = 0.2$ K (green circles). On the contrary, for $H \parallel a$, the magnetization increases monotonically until saturation, pointing to anisotropic interactions in the Hamiltonian. At $T = 0.2$ K, the plateau phase starts at the critical field $\mu_0H_{c1} \approx 7.4(2)$ T and ends at $\mu_0H_{c2} \approx 10.0(2)$ T. The saturation field is $\mu_0H_{\text{sat}} \approx 23.5(5)$ T, and the saturated moments are 4.72 $\mu_B$ when $H \parallel c$ and 4.42 $\mu_B$ when $H \parallel a$, suggesting a weak easy-axis type anisotropy. The corresponding in-plane and out-of-plane components of the g-tensor can be estimated as $g_{a} =1.77$ and $g_c =1.86$, respectively. The reduced g-factors from the free-ion value is due to crystal-field effects and spin-orbit coupling. Assuming the simplest nearest-neighbor XXZ Hamiltonian, one can use the exact spin wave result for the saturation field $g_c\mu_BH_{\text{sat}} = 3S(J_{xy}+2J_{zz})$. Because $S=5/2$ is large, we can estimate the exchange parameters from $H_{\text{sat}}$ and the classical critical field for the onset of the plateau phase $g_c\mu_BH_{\text{c1}} = 3SJ_{xy}$ \cite{Miyashita1986,Yamamoto2014} as $J_{zz} = 0.115(2)$ meV and $J_{xy} = 0.106(3)$ meV. This gives the exchange ratio $\Delta \equiv J_{xy}/J_{zz} \approx 0.92(3)$, revealing that the system is very close to the Heisenberg limit.

\subsection{Specific heat}

In order to establish the magnetic phase diagram, we have performed specific heat measurements with field applied along both the $c$ and $a$ directions. The results are plotted as $C_p/T$ in Figure~\ref{fig:specific_heat}(a) and (b), with representative temperature cuts measured in constant fields shown in  Figure~\ref{fig:specific_heat}(c) and (d). At zero field, the specific heat shows a pronounced anomaly at $T_N \approx 4$ K, indicating the development of an antiferromagnetic order, in agreement with the magnetic susceptibility data. When the field is applied along the easy axis ($H \parallel c$), the anomaly gradually shifts towards higher temperature, reaching about 4.8 K at $\mu_0H =6.5$ T, and then starts to decrease. When the field exceeds 2 T, an additional anomaly at lower temperature emerges, visible as a shoulder in the temperature dependent scans marked by the arrows; see Figure~\ref{fig:specific_heat}(c).  This feature moves towards higher fields as the temperature decreases. At $T=0.2$ K, it coincides with the onset field of the 1/3 plateau phase. For $H \parallel a$, the phase diagram is qualitatively similar, except that the variation of the ordering temperature in field is much smaller, and the lower temperature transition occurs at higher fields. 

The possible magnetic structures in the ordered phases may be inferred from theory. For the classical triangular-lattice XXZ Hamiltonian with easy-axis exchange anisotropy, early theoretical work has predicted that the ground state is ordered with a co-planar Y structure \cite{Miyashita1986}. Applying a magnetic field along the easy axis induces a sequence of phase transitions: first to the up-up-down (uud) phase, corresponding to a 1/3 magnetization plateau, then to a V phase, and finally to the fully polarized state \cite{Miyashita1986}. This sequence seems consistent with our experimental observations. The transition from the Y to the uud phase is clearly resolved in our heat capacity measurements. While the boundary between the uud and the V phase is less obvious, it can be extracted from the derivatives $dM/dH$ from the pulsed-field magnetization (red squares). Notably, the uud phase extends for a wider field range at higher temperatures, indicating that it is stabilized by thermal fluctuations as expected for large-$S$ systems \cite{Seabra2011}.

For a transverse field applied perpendicular to the easy axis, theoretical studies \cite{Yamamoto2019} have shown that it first stabilizes an inverted Y structure, with one spin pointing along the field direction and the staggered moment of the other two sublattices lying in a plane perpendicular to the field. As the field increases further, the magnetic structure becomes a $\Psi$ state, as shown in  Figure~\ref{fig:specific_heat}(b). Note that since the inverted Y and the $\Psi$ states are equivalent and continuously connected, no phase transition is expected up to the saturation. The broad heat capacity feature seen for $H \parallel a$ (dashed lines) is likely a crossover rather than a phase transition.

\subsection{Neutron diffraction}

\begin{figure}
	\includegraphics[width=\columnwidth]{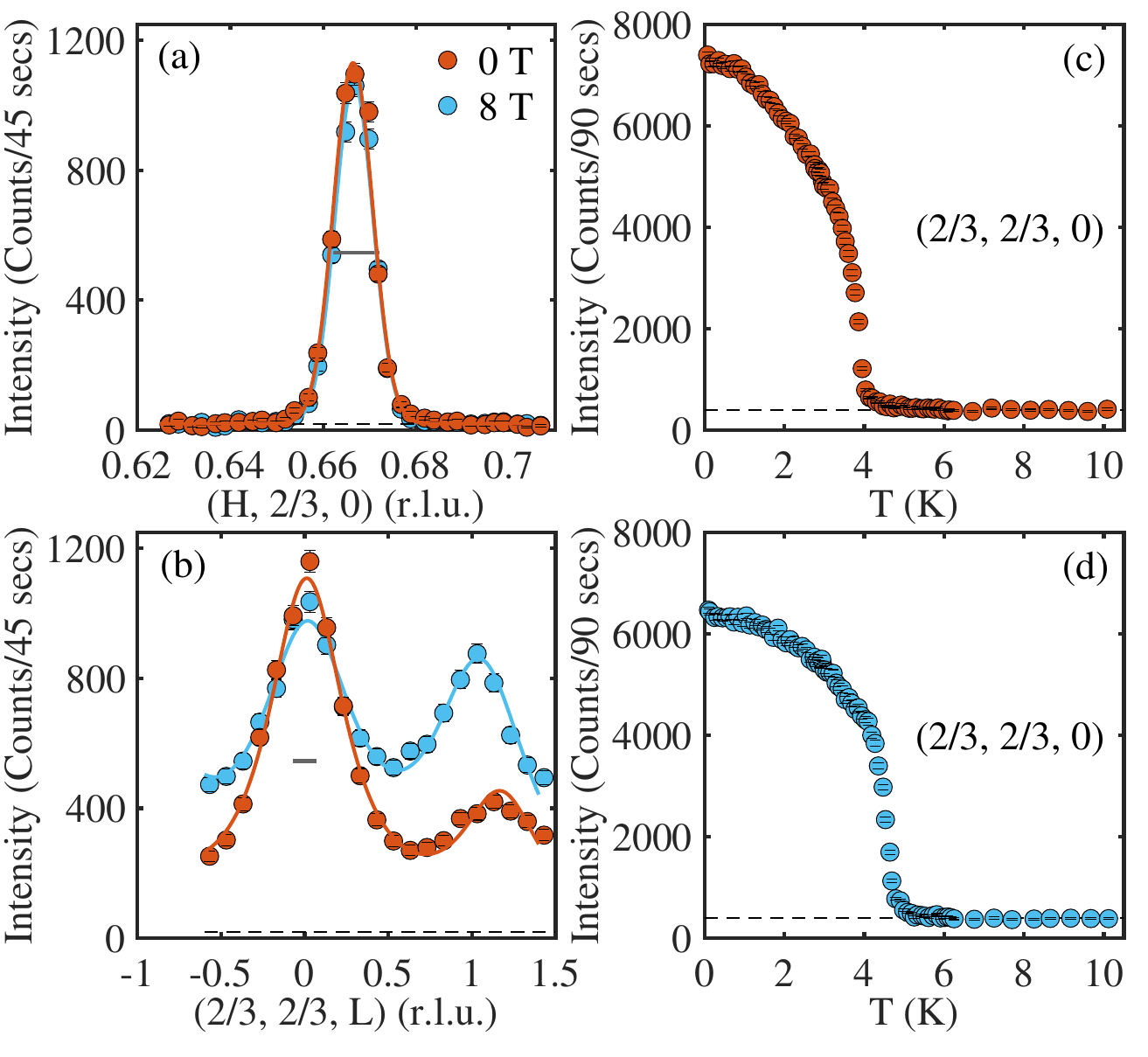}
	\caption{(a) $H$- and (b) $L$-scans over magnetic reflection $\mathbf{Q}$ = (2/3,2/3,0) measured at $\mu_0H$ = 0 and 8 T at $T=0.08$ K. Solid lines in (a) are the fits by Gaussian functions. Solid lines in (b) are the fits by Voigt functions as described in the text. Dashed lines are the background. The gray bar represents the instrumental resolution estimated using a nearby nuclear peak (1,1,0). A 40$^{\prime}$ horizontal collimator is placed after the sample to improve the Q-resolution for the $L$ scans. (c),(d) Temperature dependence of the intensity of magnetic reflection (2/3,2/3,0) measured at $\mu_0H$ = 0 and 8 T, respectively. Dashed lines are the background. No collimator is used. }\label{fig:diffraction}
\end{figure}

\begin{figure*}[th]
	\includegraphics[width=\textwidth]{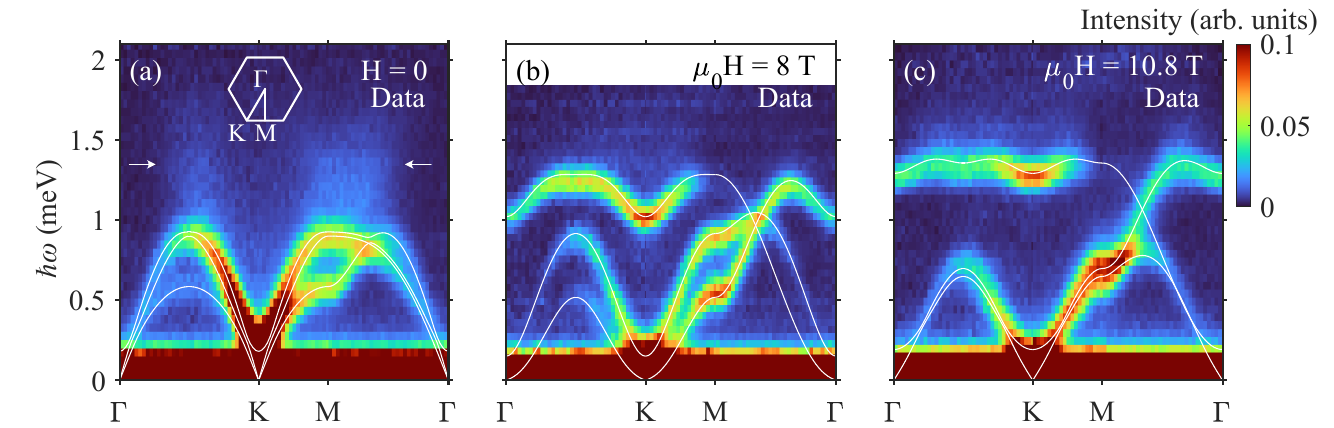}
	\caption{False color plots of magnetic excitation spectra of \KMSO measured at $T = 0.2$ K in (a) $H=0$ in the Y phase, (b) $\mu_0H = 8$ T in the uud plateau phase, and (c) $\mu_0H = 10.8$ T in the V phase, along high-symmetry directions in the reciprocal space. Solid lines are the dispersion relation calculated by LSWT using parameters summarized in Table~\ref{tab:exchange_paramters}.  White hexagon in (a) represents the boundary of the first Brillouin zone. White arrows in (a) highlight the high-energy continua. The data have been averaged over all equivalent paths as shown in Fig.~\ref{fig:reciprocal_path}, Appendix~\ref{AppendixA}.}\label{fig:INS_map}
\end{figure*}

To resolve the magnetic structures of the ordered phases experimentally, we have performed neutron diffraction measurements. In both zero field and a field of 8 T applied along the $c$ axis, magnetic reflections are observed at wave vectors $\mathbf{Q} = $(1/3,1/3,$L$) and many equivalent positions in the reciprocal space, implying a three-sublattice antiferromagnetic order consistent with the theoretically predicted Y and uud structures.  Figure~\ref{fig:diffraction}(a) and (b) show the $H$- and $L$-scans over the magnetic reflection $\mathbf{Q}$ = (2/3,2/3,0) measured at $T=0.08$ K. The $H$-scans can be well fitted using Gaussian functions, and the width is resolution limited, as indicated by the horizontal gray bar estimated using the nearby nuclear Bragg peak (1,1,0). This implies a long-range order in the triangular plane. Nevertheless, the $L$-scans are very broad and modulated in intensity, indicating a quasi-two-dimensional magnetic order similar to the isostructural \KCSO reported previously \cite{Zhu2024,Chen2024}. The intensity maxima are centered roughly at integer values of $L$, suggesting that the unit cells are likely ferromagnetically stacked along the $c$ direction. The width is much broader than the instrumental resolution (gray bar), implying that the correlation length is very short. We have fitted the $L$-scans by Voigt functions centered at $L=0$ and $\pm 1$ (solid lines), with the Gaussian width fixed by the resolution, as shown in Fig.~\ref{fig:diffraction}(b). We find that the correlation length is approximately 11 \AA~and 9 \AA~at zero field and 8 T, respectively, which amount to about half of the unit cell length. A diffuse neutron scattering experiment is necessary to fully understand the stacking and interlayer correlations between the triangular planes. 

Figure~\ref{fig:diffraction}(c) and (d) show the temperature dependence of the neutron intensity of $\mathbf{Q}$ = (2/3,2/3,0) measured at zero field and in $\mu_0H=8$ T applied along the $c$ axis. The intensity decreases gradually as temperature increases, and disappears at $T_N \approx$ 4 K at zero field and $T_N \approx$ 4.7 K at 8 T, in agreement with the specific heat anomalies shown in Figure~\ref{fig:specific_heat}(a). Unfortunately, as the magnetic ordering is quasi-2D, the Lorentz factor which associates the measured neutron intensity with the static structure factor is not well defined. Therefore, determination of the magnetic structures by refinements is not feasible. In addition, due to the short-range nature of the magnetic order, the intensity of the magnetic reflections and the modulation in the $L$ direction is dependent on the field history, which further complicate the analysis.

\subsection{Inelastic neutron scattering}


\begin{figure*}[ht]
	\includegraphics[width=\textwidth]{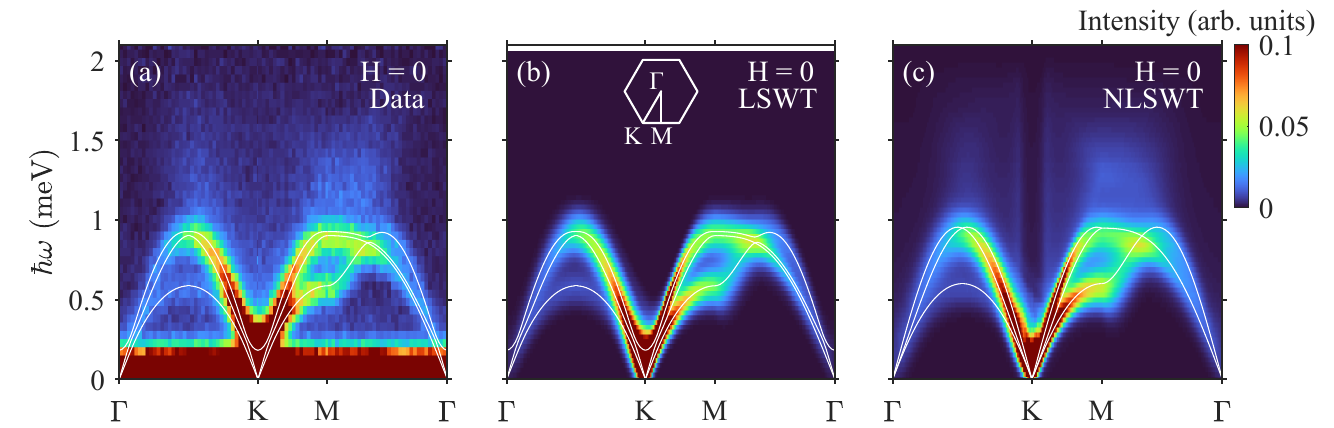}
	\caption{False color plots of magnetic excitation spectra of \KMSO at $H=0$ in the Y phase (a) measured at $T = 0.2$ K, (b) calculated by LSWT using parameters summarized in Table~\ref{tab:exchange_paramters}, and (c) calculated by non-linear SWT based on a Heisenberg model as described in the text. The data in (a) have been averaged over equivalent paths marked by stars in Fig.~\ref{fig:reciprocal_path} involving only sample rotations. The calculated intensities have been convoluted with instrumental resolution. For the non-linear SWT results in (c), the artifact in the unconvoluted spectra near the $K$ point (Fig.~\ref{fig:unconvoluted_NLSWT}, see Appendix~\ref{AppendixB}) has been masked before convolution. Solid lines are the dispersion relation of single-magnon excitations calculated by LSWT. The white hexagon in (b) represents the boundary of the first Brillouin zone.  }\label{fig:INS_map_0T}
\end{figure*}

To determine the spin Hamiltonian, we have measured the magnetic excitation spectra using inelastic neutron scattering. Figures~\ref{fig:INS_map}(a)-(c) show the false color plots of the neutron intensity as a function of energy and wave vector along high-symmetry directions of the reciprocal space, measured at $T = 0.2$ K in the zero-field Y phase, the uud phase at $\mu_0H=8$ T, and the high-field V phase at $\mu_0H=10.8$ T, respectively. To improve the counting statistics and highlight potential continuum of excitations, the intensity has been averaged over many equivalent paths as illustrated in Fig.~\ref{fig:reciprocal_path}; see Appendix~\ref{AppendixA}. The measured neutron intensities are contributed by both in-plane and out-of-plane components of the dynamical structure factor $ I \propto g_{ab}^2S^{ab}+g_c^2S^{c}$, where $S^{ab}=\frac{1}{2}(S^{xx}+S^{yy})$, $S^c = S^{zz}$, and $g_{ab}$ and $g_c$ are in-plane and out-of-plane components of the g-tensor.

In zero field shown in Figure~\ref{fig:INS_map}(a), single-magnon-like excitations are observed below 1 meV. Moreover, additional continuum  of excitations are seen at higher energy, extending up to approximately 1.8 meV. The continuum is most pronounced near the $M$ point, as well as half way between the $\Gamma$ and $K$ points, as indicated by the arrows. 

For the uud phase at $\mu_0H=8$ T shown in Figure~\ref{fig:INS_map}(b), an upper gaped mode is detected around 1.1 meV. Meanwhile, the continuum of excitations at the $M$ points have disappeared. In the V phase at $\mu_0H=10.8$ T shown in Figure~\ref{fig:INS_map}(c), the spectrum remains qualitatively similar to that in Fig.~\ref{fig:INS_map}(b), with the upper gaped mode shifted to higher energy, reaching approximately 1.35 meV while its bandwidth is reduced. The energies of the two lower branches become much closer along the $\Gamma-K-M$ path. A continuum may exist above the gaped mode in both uud and V phases, but it appears much weaker and more difficult to resolve convincingly compared with the zero-field spectrum. Solid lines in all three panels are LSWT fits discussed below.

\subsection{Theory comparison}

Now we provide a detailed comparison of the observed magnetic excitation spectra with the theory. In principle, the spin wave theory calculations should be performed based on a fully refined magnetic structure. As this is not feasible in the present case, our analysis is instead based on the simplest nearest-neighbor XXZ Hamiltonian and spin structures that are well established theoretically. These are expected to be the dominant terms in the Hamiltonian. Additional terms allowed by symmetry may be present. However, as illustrated below, the good agreement between the measured spectra and our calculation indicates that such terms are small, and are not expected to qualitative affect the main spectral features within the experimental resolution. The key features of the spectrum, such as the dispersion energies and the overall bandwidth, are primarily determined by the leading exchange interactions and are relatively less sensitive to small additional interactions or anisotropies. The lack of a fully refined magnetic structure indeed introduces some uncertainty in the extracted exchange parameters, and may obscure weak additional terms in the Hamiltonian. Therefore, our results should be understood as the minimum model that captures the leading physics, while the subleading terms require more precise information on the magnetic structure to be uniquely determined. 

To make the comparison quantitative, in Fig.~\ref{fig:INS_map_0T}(a) we show the measured spectra along a selected path (0,1,0)$-$(1/3,1/3,0)$-$(1/2,1/2,0)$-$(0,1,0) at $\mu_0H=0$ T. To improve the counting statistics without altering the instrumental resolution, only equivalent paths generated by sample rotations are averaged, as illustrated by the stars in Fig.~\ref{fig:reciprocal_path}.

\begin{table}[h!]
\caption{Exchange parameters and components of $g$-tensor extracted using the magnetization data and by fitting the inelastic neutron scattering spectra using LSWT, as well as the Heisenberg exchange $J$ employed in the NLSWT.}
\label{tab:exchange_paramters}
\begin{ruledtabular}
\begin{tabular}{ccccc}
 & $J_{zz}$ (meV) & $\Delta \equiv J_{xy}/J_{zz}$ & $g_c$ & $g_{ab}$ \\ \hline
Magnetization & 0.115(2) & 0.92(3) & 1.86 & 1.77 \\ 
INS, 0 T - LSWT & 0.117(2) & 0.97(3) & $-$ & $-$ \\ 
INS, 8 T - LSWT & 0.135(2) & 0.96(1) & 2.06(2) & $-$ \\ 
INS, 10.8 T - LSWT & 0.126(1) & 0.92(1) & 2.04(2) & $-$ \\
INS, 0 T - NLSWT & 0.120 & 1 & $-$ & $-$
\end{tabular}
\end{ruledtabular}
\end{table}

\begin{figure*}[th]
	\includegraphics[width=\textwidth]{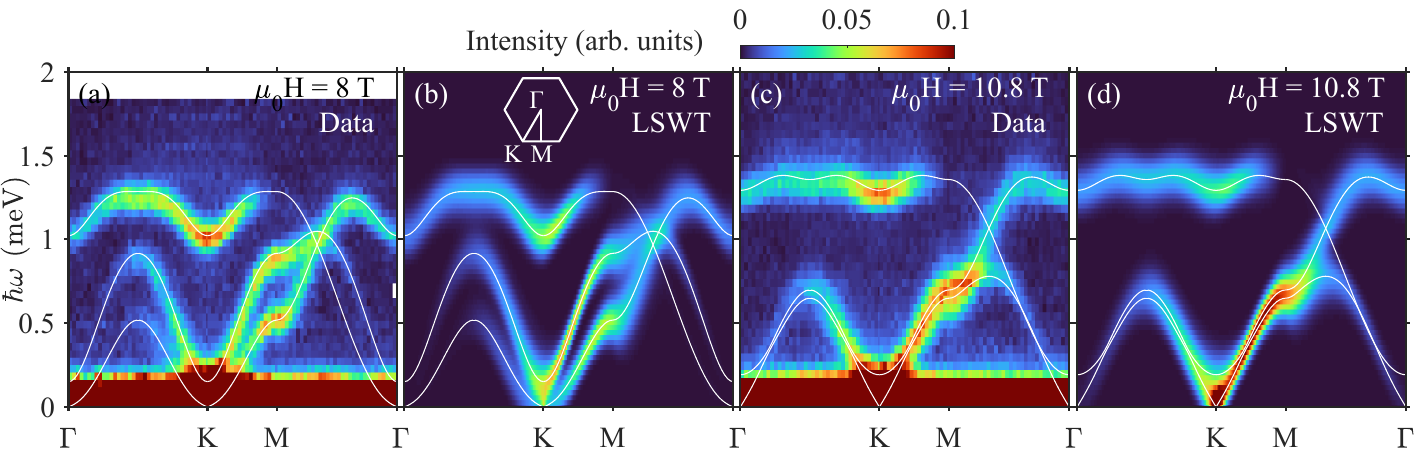}
	\caption{False color plots of magnetic excitation spectra of \KMSO measured at $T = 0.2$ K at (a) $\mu_0H = 8$ T in the uud phase, and (c) $\mu_0H = 10.8$ T in the V phase. The data have been averaged over equivalent paths marked by stars in Fig.~\ref{fig:reciprocal_path} involving only sample rotations. (b) and (d) are the corresponding spectra calculated by LSWT using parameters summarized in Table~\ref{tab:exchange_paramters}. The calculated intensities have been convoluted with instrumental resolution. Solid lines denote the calculated dispersion relation of single-magnon excitations. The white hexagon in (b) represents boundary of the first Brillouin zone.  }\label{fig:INS_map_8T_10p8T}
\end{figure*}

The corresponding zero-field spectrum calculated within LSWT based on the XXZ Hamiltonian and Y spin structure is shown in Fig.~\ref{fig:INS_map_0T}(b), where the intensity has been convoluted with the instrumental resolution. We find that the single-magnon excitations below 1 meV can be well reproduced using the exchange parameters in Table~\ref{tab:exchange_paramters}. However, the LSWT cannot capture the high-energy continuum, which underscores the central role played by magnon-magnon interactions owing to the non-collinear spin structure \cite{ZhitomirskyChernyshev_RMP_2013_DecayReview}.

To take into account the effects of magnon-magnon interactions, we performed non-linear SWT (NLSWT) simulations. As the exchange anisotropy is weak, its primary effects are limited to a small gap opening in one magnon branch and slight shifts in magnon energies, as illustrated by the solid lines in Fig.~\ref{fig:INS_map_0T}(b) and (c). Given the coarse experimental resolution ($\Delta E \approx 0.19$ meV), these subtle features are difficult to be resolved reliably in the present experiment. Future high-resolution inelastic neutron scattering measurements are necessary to resolve the effects of such weak exchange anisotropy. In light of this, our NLSWT calculation was carried out based on a pure Heisenberg model with $S=5/2$, following previous work \cite{Mourigal2013_TAL}. The Heisenberg exchange $J$ was set to be $J = 0.120$ meV to better reproduce the observed magnon energies. The resulting resolution-convoluted zero-field excitation spectra are shown in Fig.~\ref{fig:INS_map_0T}(c). The agreement with the experimental data is rather striking, with both single-magnon excitations and high-energy continuum accurately reproduced. For completeness, the unconvoluted NLSWT excitation spectrum, also showcasing the contribution from different polarization channels is shown in Appendix.~\ref{AppendixB},  Fig.~\ref{fig:unconvoluted_NLSWT}.

The effects that are beyond LSWT are threefold. First is the decay and renormalization of magnons due to magnon interactions. Second is the spectral weight transfer from one sector to the other, for example, from the one-magnon branch to the two-magnon continuum. Third is not specific to the magnon interaction as a source, but is the direct contribution of the two-magnon continuum through the longitudinal component of the dynamical structure factor \cite{Mourigal2013_TAL}. 
The decay of magnons leads to broadening of their linewidth. Our non-linear SWT calculation indicates that, after singularities are regularized, such broadening is approximately independent of the spin value, with a maximum broadened width of about $0.2J$ \cite{Chernyshev2009}. This corresponds to roughly 0.024 meV, far below the experimental energy resolution of about 0.19 meV. Thus, direct verification of decay-induced broadening is not possible in the present experiment.
However, the second and third effect are clearly detectable, together with the renormalization discussed above. The spectral weight transfer manifests itself in the transverse component of the dynamical structure factor, inheriting coupling to the two-magnon continuum and the Van Hove singularities associated with it \cite{Chernyshev2006,Chernyshev2009,Mourigal2013_TAL}, as shown in  Fig.~\ref{fig:unconvoluted_NLSWT}(b) and (c) in Appendix~\ref{AppendixB}. This, together with the longitudinal two-magnon scattering, gives rise to the continuum above the single-magnon excitations.

It is worth noting that the exchange interaction $J$ extracted from NLSWT is slightly larger than that obtained from LSWT fit in Fig.~\ref{fig:INS_map_0T}(b). Such a renormalization is expected within the $1/S$ paradigm. In this framework, there are two sources leading to the renormalization: the quartic, i.e., mean-field or Hartree-Fock terms, and the cubic terms that couple one- and two-magnon sectors. They can be of opposite signs \cite{ZhitomirskyChernyshev_RMP_2013_DecayReview}. In the zero-field Y phase, both terms are present. The net effect is a small downward shift of the magnon energies in the non-linar SWT calculations relative to LSWT, or equivalently, an upward renormalization of $J$ for the NLSWT by approximately 2.6\%, as shown in Table~\ref{tab:exchange_paramters}.

\begin{figure*}[ht]
	\includegraphics[width=\textwidth]{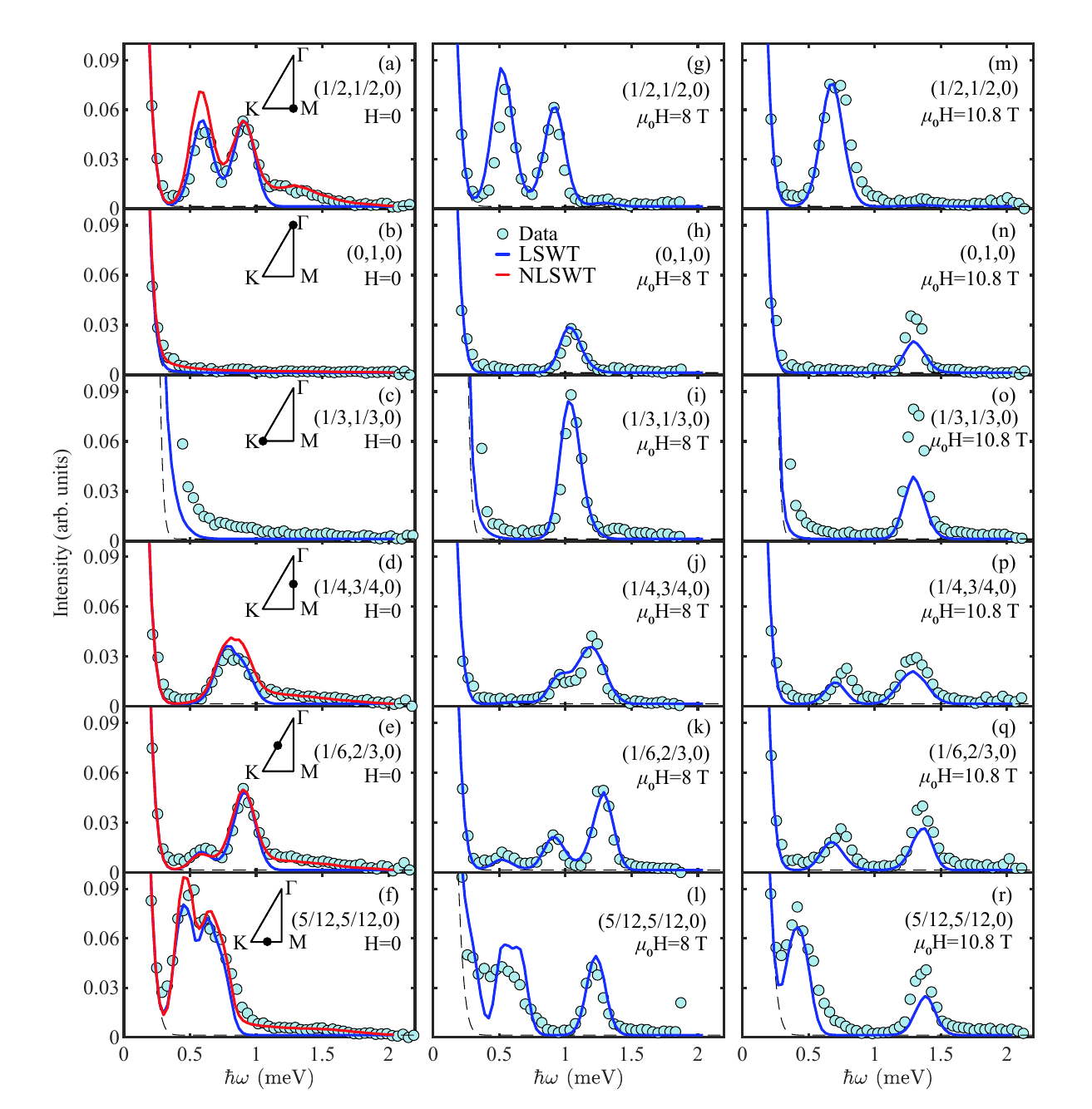}
	\caption{Constant-$\mathbf{Q}$ cuts of the magnetic excitation spectra of \KMSO at representative wave vectors, measured at $T = 0.2$ K in (a)-(f) $H$ = 0, (g)-(l) $\mu_0H$ = 8 T, and (m)-(r) $\mu_0H$ = 10.8 T. The data are obtained by integrating over $\pm$0.025~\AA$^{-1}$ along two orthogonal directions in the reciprocal space around the selected wave vector. The error bars are smaller than the symbol size. Blue and red solid lines are calculated curves using LSWT and non-linear SWT, respectively, and have been convoluted with the experimental resolution and averaged over the region of integration. An arbitrary scale factor has been applied to match the intensity at the $M$ point. Dashed lines are fits of the elastic peak and a constant background. }\label{fig:INS_cut}
\end{figure*}

Similarly, Fig.~\ref{fig:INS_map_8T_10p8T}(a) and (c) show the magnetic excitation spectra for 8 T and 10.8 T in the uud and V phases, respectively, along the selected paths described above. The corresponding LSWT simulations are plotted in (b) and (d). The best agreement between the data and simulation is achieved using the exchange parameters summarized in Table~\ref{tab:exchange_paramters}. Notably, the exchange parameters and the g-tensor are renormalized relative to the values extracted from the magnetization data. The magnitude of renormalization in $J_{zz}$ ($J_{xy}$) is approximately 17\% (22\%) in the uud phase and 10\% (15\%) in the V phase, compared to 2\% (7\%) in the zero-field Y phase. The renormalization in $g_c$ is about 10\% in both fields. We note that the extracted exchange constants and g-factors should be regarded as effective parameters, as LSWT is strictly exact only in the field-polarized state. Nevertheless, they provide a quantitative estimate of the magnitude of quantum effects in this $S=5/2$ system. Overall, the renormalization is moderate and consistent with the expectation of reduced, yet still non-negligible quantum fluctuations in a frustrated large-$S$ system.

At present,  NLSWT calculations for the uud or V phases are not available. Nevertheless, qualitative trends can be inferred on general grounds. The collinear uud phase lacks cubic interaction terms, hence there is likely stronger upward renormalization of $J$ for the LSWT fits due to quartic interactions. A continuum may still arise from these quartic terms as well as from the two-magnon scattering. In contrast, both cubic and quartic terms are at play in the V phase. In addition, the V phase lies closer to the saturation field $H_{\text{sat}}$, above which LSWT is exact. Therefore, $J$ extracted from the LSWT fit in this phase should be closer to the ``bare" $J$ in the Hamiltonian. These trends in the uud and V phases also broadly correspond to the lesser visibility and hence lesser role of the two-magnon continuum.

In Fig.~\ref{fig:INS_cut} we show the constant-$\mathbf{Q}$ cuts of the experimental data at representative wave vectors, along with the calculated curves from linear (blue) and non-linear SWT (red). In zero field, shown in Fig.~\ref{fig:INS_cut}(a)-(f), the improved agreement of non-linear SWT over LSWT is evident, particularly in reproducing the high-energy tail arising from the continuum. Notably, at the $M$ point shown in Fig.~\ref{fig:INS_cut}(a), a small peak appears near 1.3 meV, which originates from the Van Hove singularity of the continuum. In addition, the extended high-energy tail also contains a direct contribution of the two-magnon continuum.

The constant-$\mathbf{Q}$ cuts of the spectra measured at 8 T and 10 T are presented in Fig.~\ref{fig:INS_cut}(g)-(r). With renormalized exchange parameters and g-tensor, LSWT reproduces the magnon energies reasonably well in both the uud phase and the V phase. Nevertheless, noticeable discrepancies remain between the measured and calculated neutron intensities, because additional effects, such as spectral weight transfer and two-magnon continuum, are not accounted for within LSWT.

\section{Discussion}

We have established that \KMSO is a model triangular-lattice antiferromagnet that is well described by a simple $S=5/2$ XXZ easy-axis Hamiltonian very close to the Heisenberg limit. Quantum effects are still observable in the magnetic excitation spectrum of such a large-spin system, and can be quantitatively understood. 

We first compare it with the $S=1/2$ counterpart \BCSO.  In both materials, the zero-field ground state has a non-collinear (nearly) 120$^{\circ}$ spin structure. 
However, their excitation spectra are qualitatively different. Although single-magnon modes and broad continuum of excitations are present in both \BCSO \cite{Itoh2017,MacdougalWilliams_PRB_2020_BaCoSbOtriangularexcitations} and \KMSO, in \BCSO they deviate strongly from the predictions of non-linear SWT \cite{Mourigal2013_TAL}. This indicates that magnon-magnon interactions in \BCSO are strong, preventing a direct quantitative comparison with the quasiclassical theory  and calling for alternative description, such as the ones utilizing Schwinger bosons or spinons \cite{Ghioldi2022,Drescher2023,Bose2025}.
In stark contrast, such a comparison for the $S=5/2$ compound \KMSO is not only fruitful and quantitative, but also clarifies distinct roles of various magnon-magnon interaction components.

Finally, we compare \KMSO with other large-$S$ triangular lattice compounds. 
To date, inelastic neutron scattering studies on single crystal samples of such systems remain scarce, with Ba$_3$MnSb$_2$O$_9$ being one of the few reported examples \cite{Mingfang2023}. This compound exhibits an easy-plane anisotropy. However, there exists a slight monoclinic structural distortion that allows for biaxial anisotropy, resulting in additional terms into the Hamiltonian. The measured excitation spectrum can be qualitatively described by LSWT. Neverthelss, the magnon linewidth observed in Ref.~\cite{Mingfang2023} is substantially broader than the instrumental resolution. This behavior stands in contrast to \KMSO. In our case, no additional terms in the Hamiltonian have been detected beyond the simple XXZ model. The linewidth of single magnon modes remains comparable to the resolution. Moreover, in \KMSO a high-energy continuum can be unambiguously resolved, whereas the energy range covered in Ref.~\cite{Mingfang2023} for Ba$_3$MnSb$_2$O$_9$ is relatively limited, preventing direct access to such features. In addition, the energy scale of the dominant exchange interaction in Ba$_3$MnSb$_2$O$_9$ is more than twice that of \KMSO studied in this work, rendering studies of the field-induced phases by neutron scattering more challenging. 


\section{Conclusion}
Calorimetric, magnetic, and neutron scattering measurements establish that the $S=5/2$ triangular lattice antiferromagnet \KMSO is well described by an XXZ easy-axis Hamiltonian close to the Heisenberg limit. The availability of large single crystals, combined with low energy scale of the exchange interactions, provides an excellent platform to study the thermodynamic and dynamical properties of all ordered phases prior to saturation. The results demonstrate that magnon-magnon interactions still play a central role in the spin dynamics of large-$S$ Heisenberg antiferromagnets on a triangular lattice, and enable tests of  interacting magnon theories on a quantitative level.

\begin{acknowledgements}
Work at ETHZ was supported by a MINT grant of the Swiss National Science Foundation. 
We acknowledge the support of the HLD at HZDR, member of the European Magnetic Field Laboratory (EMFL), and the W\"urzburg-Dresden Cluster of Excellence on Complexity, Topology and Dynamics in Quantum Matter--$ctd.qmat$ (EXC 2147, Project No.\ 390858490). This work is based on experiments performed at the Swiss spallation neutron source SINQ, Paul Scherrer Institute, Villigen, Switzerland.
The work of A.L.C. on the NLSWT was supported by the U.S. Department of Energy, Office of Science, Basic Energy Sciences under Award No. DE-SC0021221.
\end{acknowledgements}

\appendix
\section*{Appendix}
\setcounter{section}{0}
\renewcommand{\thesection}{\Alph{section}}
\renewcommand{\thefigure}{A\arabic{figure}}
\setcounter{figure}{0}
\renewcommand{\theequation}{\arabic{equation}}
\setcounter{equation}{0}

\section{Averaging of INS data over equivalent paths in the reciprocal space}\label{AppendixA}

\begin{figure*}[ht]
	\includegraphics[width=0.9\textwidth]{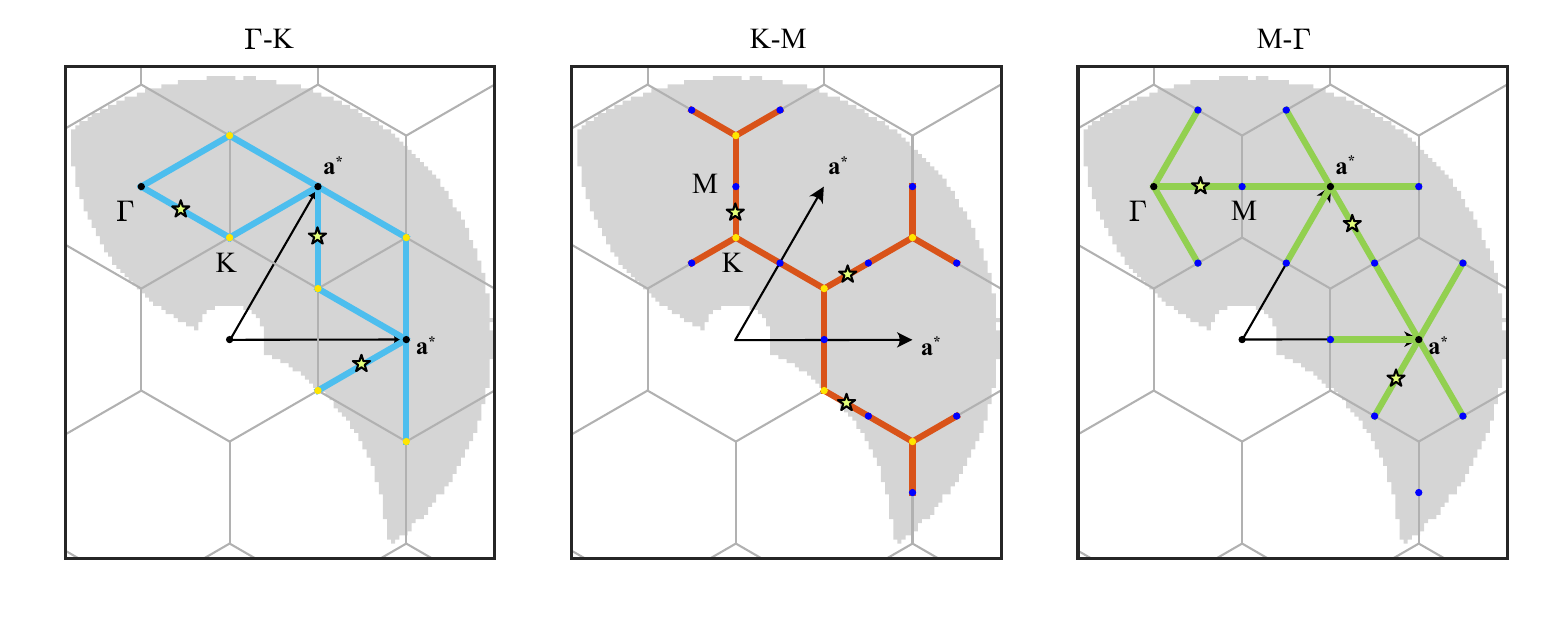}
	\caption{Schematic diagrams illustrating the detector coverage of CAMEA at elastic scattering and the equivalent reciprocal paths along (a) $\Gamma$-K, (b) K-M, and (c) M-$\Gamma$ directions used to generate Fig.~\ref{fig:INS_map}. The hexagons represent Brillouin zone boundaries. The stars mark the paths used to produce Fig.~\ref{fig:INS_map_0T}(a) and Fig.~\ref{fig:INS_map_8T_10p8T}(a),(c), which are related to one another by sample rotations only. }\label{fig:reciprocal_path}
\end{figure*}

Fig.~\ref{fig:reciprocal_path} illustrates the equivalent reciprocal paths used to plot the averaged excitation spectra in Fig.~\ref{fig:INS_map}, Fig.~\ref{fig:INS_map_0T}(a) and  Fig.~\ref{fig:INS_map_8T_10p8T}(a),(c).

\renewcommand{\thefigure}{B\arabic{figure}}
\setcounter{figure}{0}
\setcounter{equation}{0}

\section{Dynamical structure factor of Heisenberg triangular-lattice antiferromagnet}\label{AppendixB}

The dynamical structure factor is related to the dynamical spin-spin correlator ${\cal S}^{\alpha\beta}({\bf q},\omega)$ via  standard multiplicative kinematic formfactors and components of the $g$-tensor. 
The non-linear spin-wave calculation of the dynamical spin correlator for the nearest-neighbor, spin $S$, Heisenberg triangular-lattice antiferromagnet  follow Ref.~\cite{Mourigal2013_TAL}. Here, we briefly recite the essential steps. 

We assume the $120^{\circ}$ spin structure to be in the $xz$ plane. The spin-wave calculations require a transformation to the rotating frame with the spin-quantization axis oriented along the local magnetization on each site. Using the ordering vector of the $120^{\circ}$  structure ${\bf Q}=(4\pi/3a,0)$,  the dynamical spin correlator ${\cal S}^{\alpha\alpha}({\bf q},\omega)$ in this rotated frame naturally splits into three components, the two transverse ($xx$ and $yy$) and the longitudinal ($zz$) contributions
\begin{eqnarray}
{\cal S}^{\perp,yy}({\bf q},\omega) &=&  {\cal S}^{yy}({\bf q},\omega), \nonumber\\
  {\cal S}^{\perp,xx}({\bf q},\omega)&=&
 \frac{1}{2}\Big( {\cal S}^{xx}({\bf q_+},\omega) + {\cal S}^{xx}({\bf q_-},\omega) \Big) ,
 \label{Sqw} \\
{\cal S}^{L}({\bf q},\omega) &=& \frac{1}{2}
\Big( {\cal S}^{zz}({\bf q_+},\omega) + {\cal S}^{zz}({\bf q_-},\omega)\Big) ,	
\nonumber
\end{eqnarray}
with  shorthand notations  ${\bf q}_\pm\!=\!{\bf q\pm Q}$ and ignoring mixed $xz$ and $zx$ fluctuations. Here $\{x,y,z\}$ refer to the local, i.e., rotated spin axes. 

The dynamical spin correlator is evaluated using Holstein-Primakoff representation of spin operators  in terms of bosons and  subsequent  Bogolyubov transformation; see Ref.~\cite{Chernyshev2009} for details.  As a result, the ${\cal S}^{xx}({\bf q},\omega)$ and ${\cal S}^{yy}({\bf q},\omega)$ contributions to Eq.~(\ref{Sqw}) can be written as 
\begin{eqnarray}
{\cal S}^{xx}({\bf q},\omega) & = & {\cal F}^{xx}_{\bf q} A({\bf q},\omega) ,\quad 
{\cal S}^{yy}({\bf q},\omega)  =  {\cal F}^{yy}_{\bf q} 
A({\bf q},\omega), \ \ \ 
\label{G_trans_1}
 \end{eqnarray}
where ${\cal F}^{xx(yy)}_{\bf q}$ contain Bogolyubov transformation parameters and Hartree-Fock averages of the  Holstein-Primakoff bosons (see Ref.~\cite{Mourigal2013_TAL} for explicit expressions), and  $A({\bf q},\omega)\!=\!-(1/\pi)\textrm{Im}\bigl[G({\bf q},\omega)\bigr]$ is the  spectral function of the single-magnon Green's function $G({\bf q},\omega)$. 

Similarly, the longitudinal ${\cal S}^{zz}({\bf q},\omega)$ contributions are
\begin{equation}
{\cal S}^{zz} ({\bf q},\omega)  =
\frac{1}{2}\! \sum_{\bf k} {\cal F}^{zz}_{\bf k,q}\delta(\omega-\varepsilon_{\bf k} - \varepsilon_{\bf k-q}),
\label{Szzqw}
\end{equation}
where ${\cal F}^{zz}_{\bf k,q}$ is another combination of the  Bogolyubov parameters~\cite{Mourigal2013_TAL} and  $\varepsilon_{\bf k}$ is the LSWT magnon energy.

The single-magnon Green's function is given by
\begin{eqnarray}
 G({\bf q},\omega) & \approx & \bigl[\omega-\varepsilon_{\bf q}-\Sigma({\bf q},\omega)\bigr]^{-1} \ ,
  \label{BE}
\end{eqnarray}
with the lowest  order magnon self-energy
\begin{eqnarray}
\Sigma({\bf q},\omega) & = &  \Sigma^{\rm HF}({\bf q}) + \Sigma^d({\bf q},\omega)  + \Sigma^s({\bf q},\omega) \ ,
\label{Sigma}
\end{eqnarray}
where $\Sigma^{\rm HF}({\bf q})$ is the frequency-independent Hartree-Fock contribution from the quartic terms of the Holstein-Primakoff expansion, and $\Sigma^d({\bf q},\omega)$  and $\Sigma^s({\bf q},\omega)$ are the three-magnon  decay and source  terms, respectively~\cite{Chernyshev2009}.

The  ``off-shell''  consideration of the spectral properties within the NLSWT often leads to spurious divergences. One of them is dealt with by keeping the formally non-singular source term of the self-energy on-shell, i.e., putting $\omega\!=\!\varepsilon_{\bf q}$ in this term \cite{Mourigal2013_TAL}.  Another common problem is numerically not exact cancellation of formally mutually canceling terms in the self-energy near the Goldstone mode, ${\bf q}\rightarrow K(={\bf Q})$, which is amplified by a divergent $1/|{\bf q}-{\bf Q}|$ combination of the Bogolyubov parameters in one of the transverse components, ${\cal S}^{yy}({\bf q},\omega)$. This artificial divergence can be suppressed by increasing the number of integration points in evaluating separate terms in $\Sigma({\bf q},\omega)$. Since, in practice, suppressing it becomes numerically costly and concerns only a small ${\bf q}$-region near the K-point, we proceed with a simple amputation procedure for this component of ${\cal S}^{yy}({\bf q},\omega)$.

\begin{figure*}[ht]
	\includegraphics[width=\textwidth]{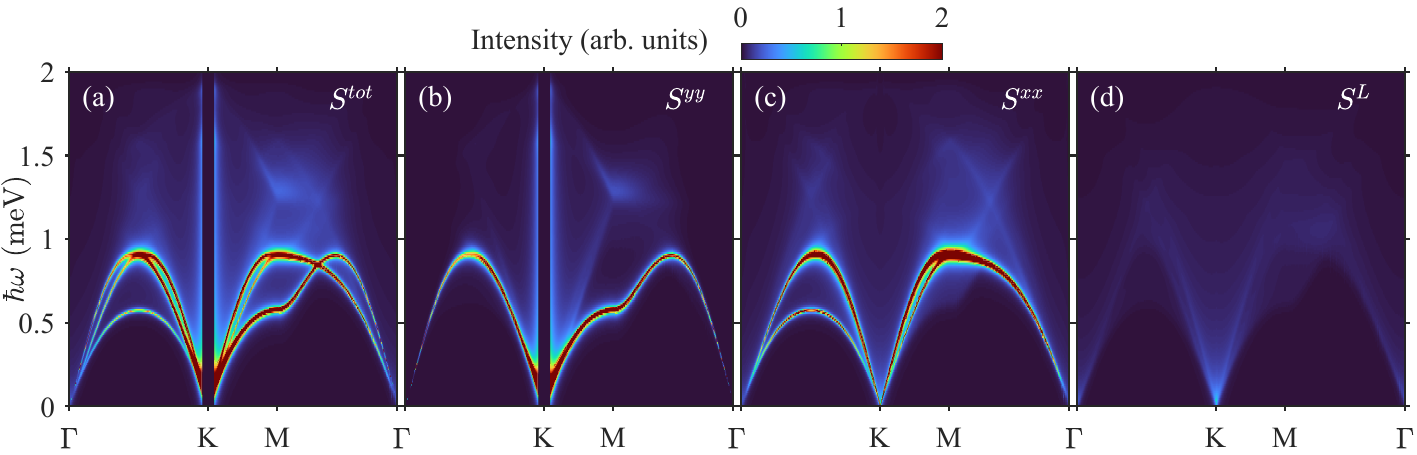}
	\caption{False color plots of the magnetic excitation spectra calculated by non-linear SWT, using a Heisenberg model as described in the text, along high-symmetry directions in the reciprocal space. The intensity has not been convoluted with the experimental resolution. $S^{tot}$, $S^{xx}$, $S^{yy}$, and $S^{L}$ denote (a) total dynamical structure factor and (b)-(d) contributions from different polarization channels in Eq.~(\ref{Sqw}). $S^{L}$ is the longitudinal two-magnon continuum. $S^{xx}$ and $S^{yy}$ are two transverse components from (\ref{Sqw}), receptively. The artifact near the $K$ point has been amputated.  }\label{fig:unconvoluted_NLSWT}
\end{figure*}

The resultant magnetic excitation spectrum of the pure $S=5/2$ Heisenberg model calculated using non-linear SWT is shown in Fig.~\ref{fig:unconvoluted_NLSWT}(a). The contributions of different polarization channels from Eq.~(\ref{Sqw}) to the total dynamical structure factor are shown in Fig.~\ref{fig:unconvoluted_NLSWT}(b)-(d). $S^L$ is the longitudinal component, whereas $S^{xx}$ and $S^{yy}$ are the two transverse components, respectively. No convolution with the instrumental resolution has been performed, while a small artificial Lorentzian broadening of $\delta=0.036$ meV has been used in Eqs.~(\ref{Szzqw}) and (\ref{BE}).

\clearpage

\bibliography{KMSO}

\end{document}